\documentclass[11pt]{article}
\usepackage{jvaliviita_blois,epsfig}
\usepackage{amsmath}
\usepackage{color}

\def\Journal#1#2#3#4{{#1} {\bf #2}, #3 (#4)}

\def\PRL{\em Phys. Rev. Lett.}
\def\PRD{{\em Phys. Rev.} D}

\def\be{\begin{equation}}
\def\ee{\end{equation}}
\def\bea{\begin{eqnarray}}
\def\eea{\end{eqnarray}}

\begin{document}

\newcommand{\nc}{\newcommand}
\nc{\half}{\textstyle\frac{1}{2}}
\nc{\fsky}{f_{\rm sky}}
\nc{\fwhm}{\theta_{\rm fwhm}}
\nc{\fwhmc}{\theta_{{\rm fwhm},c}}
\nc{\nadi}{n_{\rm ad1}}
\nc{\nadis}{n_{\rm ad2}}
\nc{\niso}{n_{\rm iso}}
\nc{\ncor}{n_{\rm cor}}
\nc{\fiso}{f_{\rm iso}}
\nc{\wmap}{\textsc{wmap} }
\nc{\camb}{\textsc{camb} }
\nc{\R}{{\cal{R}}}

\vspace*{4cm}
\title{CORRELATED ADIABATIC AND ISOCURVATURE CMB FLUCTUATIONS IN THE LIGHT
OF THE WMAP DATA
}

\author{ J. V\"ALIVIITA }

\address{Department of Physical Sciences, University of Helsinki, and
  Helsinki Institute of Physics,\\
   P.O. Box 64, FIN-00014 University of Helsinki, Finland}

\maketitle\abstracts{%
  In multi-field inflation models, correlated adiabatic and isocurvature
  fluctuations are produced and in addition to the usual adiabatic fluctuation
  with a spectral index $\nadi$
  there is another adiabatic component with a spectral index
  $\nadis$ generated by entropy perturbation during inflation,
  if the trajectory in the field space is curved.
  Allowing $\nadi$ and $\nadis$ to vary independently
  we find that the \wmap data favor models where the two adiabatic components
  have opposite spectral tilts. This leads naturally to a running adiabatic
  spectral index. The \wmap data  with a prior $\niso < 1.84$ for the isocurvature
  spectral index gives $\fiso < 0.84$ for the isocurvature fraction of the
  initial power spectrum at $k_0=0.05$ Mpc$^{-1}$. We also comment on a
  degeneration between the correlation component and the optical depth
  $\tau$. Moreover, the measured low quadrupole in the TT angular power
  could be achieved by a strong negative correlation, but then one would need
  a large $\tau$ to fit the TE spectrum.%
}

\section{Introduction}
This talk given in the ``15$^{\rm th}$ Rencontres de Blois'' is partially based on
a Letter by me and V.\ Mu\-ho\-nen \cite{Valiviita:2003ty}. I consider a
correlation between adiabatic and cold dark matter (CDM) isocurvature initial
perturbations in more realistic models than the \wmap group did in \cite{Peiris:2003ff}.
Prior to the \wmap
it was not possible to make reasonable constraints on the correlated
models, but now the accurate enough TT (temperature-temperature, i.e.
temperature auto correlation, $C_l^{TT}$) \cite{Bennett:2003bz}
and TE (temperature-polarization E-mode cross-correlation,  $C_l^{TE}$) \cite{Kogut:2003et}
power spectra are available.


In Fig.~1(a) I show schematically the evolution of the universe.
The horizontal axis is the scale factor and the vertical axis
the physical length scale.
During inflation there are quantum fluctuations that freeze in when a
particular scale goes out of the horizon.
This figure is to explain two instants of time appearing 
later in my talk. The horizon exit of
cosmologically interesting scales during inflation is marked by
$t_*$. However, the initial conditions for the CMB angular power
calculations (e.g., for \camb code \cite{jvaliviita_camb}) should be given deep in the
radiation dominated era. Thus the interesting ``initial time''
for CMB physicists is $t_{\rm rad}$.
\begin{figure}
\begin{center}
{\normalsize\bf (a)}
{\tiny
\setlength{\unitlength}{0.35703mm}
\begin{picture}(170,120)(0,0)
\put(0,0){\resizebox{60.690mm}{!}{
\includegraphics{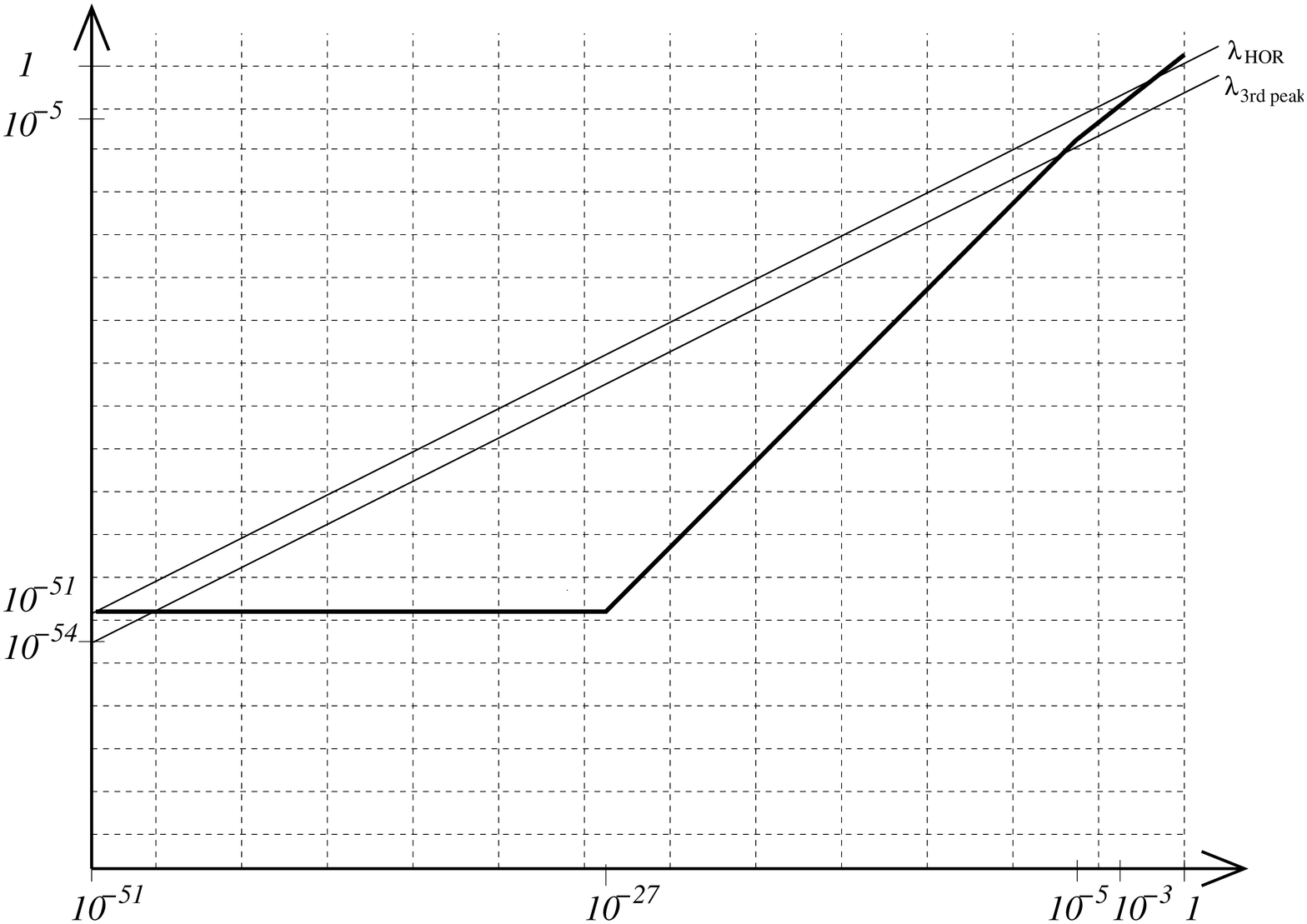}}}
\put(35,18){\textcolor{magenta}{INFLATION}}
\put(35,27){\shortstack{
\textcolor{magenta}{horizon $H^{-1}$}\\ 
\textcolor{magenta}{$\approx$ constant}}} 
\put(90,13){\shortstack{\textcolor{red}{RADIATION}\\ 
\textcolor{red}{DOM.}}}
\put(140,13){\shortstack{\textcolor{blue}{MATTER}
\\ \textcolor{blue}{DOM.}}}
\put(115,70){\textcolor{red}{$H^{-1}\propto a^2$}}
\put(140,90){\textcolor{blue}{$H^{-1}\propto a^{3/2}$}}
\thicklines
\put(110,60){\frame{HORIZON}}
\put(110,60){\vector(-1,0){10}}
\put(90,40){\frame{\shortstack[l]{end of inflation,\\reheating}}}
\thicklines
\put(90,40){\vector(-1,0){7}}
\put(100,103){\frame{\shortstack[l]{$\rho_r=\rho_m$,\\$a\sim10^{-5}$}}}
\put(100,103){\vector(1,0){40}}
\put(20,95){\frame{\shortstack[l]{physical size of\\ todays horizon scale}}}
\put(20,95){\vector(1,0){100}}
\put(20,75){\frame{\shortstack[l]{physical size of\\ the scale of
the\\ 3rd acoustic peak}}}
\put(20,75){\vector(1,0){69}}
\put(20,55){\frame{\shortstack[l]{$\lambda_{HOR} \approx$\\ $10^{-26}$m}}}
\put(20,55){\vector(-1,-2){7}}
\put(16,9){\frame{\shortstack[l]{q\\u\\a\\n\\t\\u\\m}
\shortstack[l]{f\\l\\u\\c\\t.\\$\phantom{a}$\\$\phantom{a}$}}}
\put(15,114){phys.\ length/$2H_0^{-1}$}
{\normalsize
\put(162,0){$a$}
\put(24,-3){$t_*$}
\put(87,-3){$t_{\rm rad}$}}
\end{picture}
\hspace{0.1cm}
{\normalsize\bf (b)}
\includegraphics[width=0.2\textwidth,height=4.3cm]{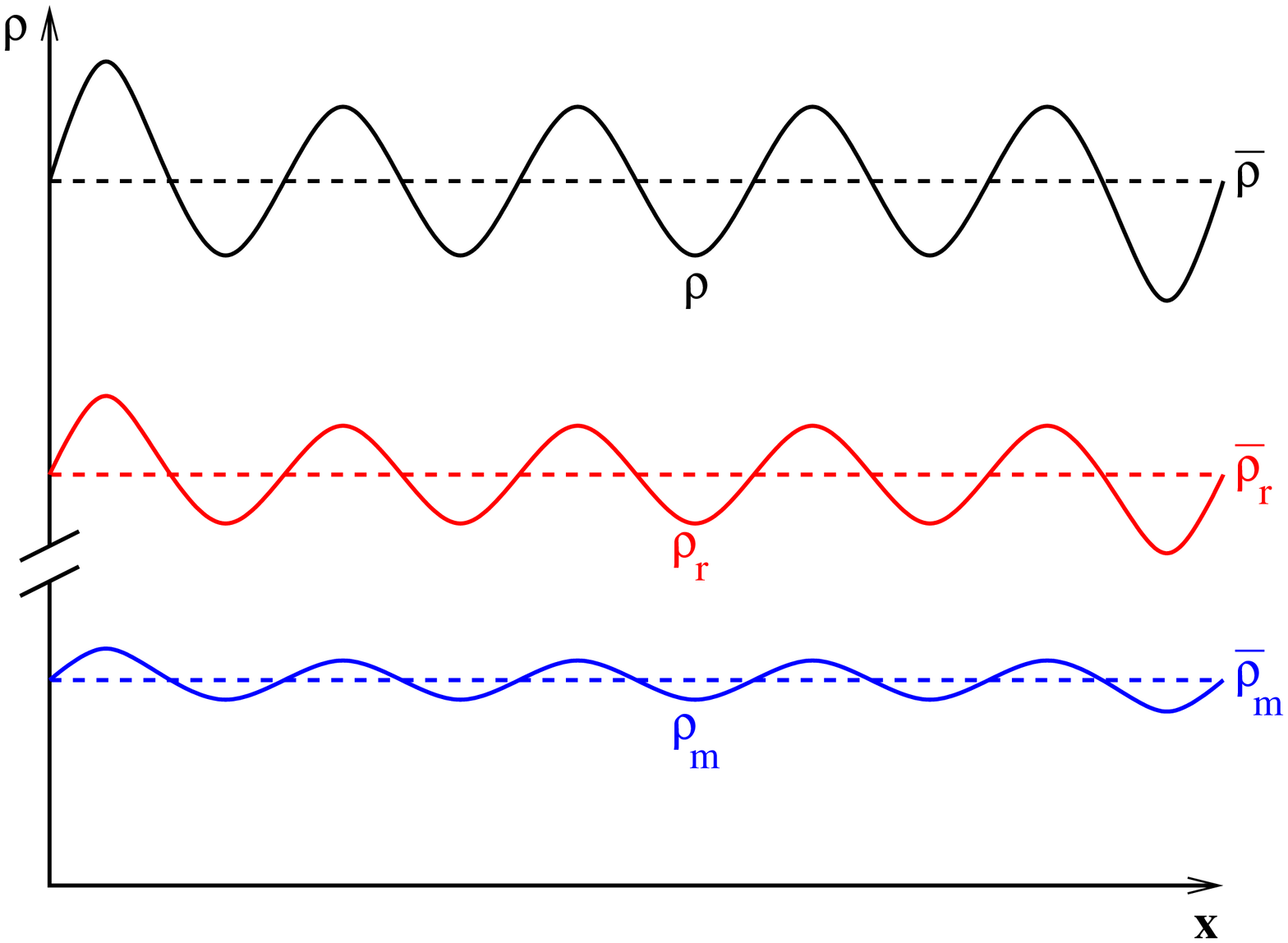}
\hspace{0.4cm}
{\normalsize\bf (c)}
\includegraphics[width=0.2\textwidth,height=4.3cm]{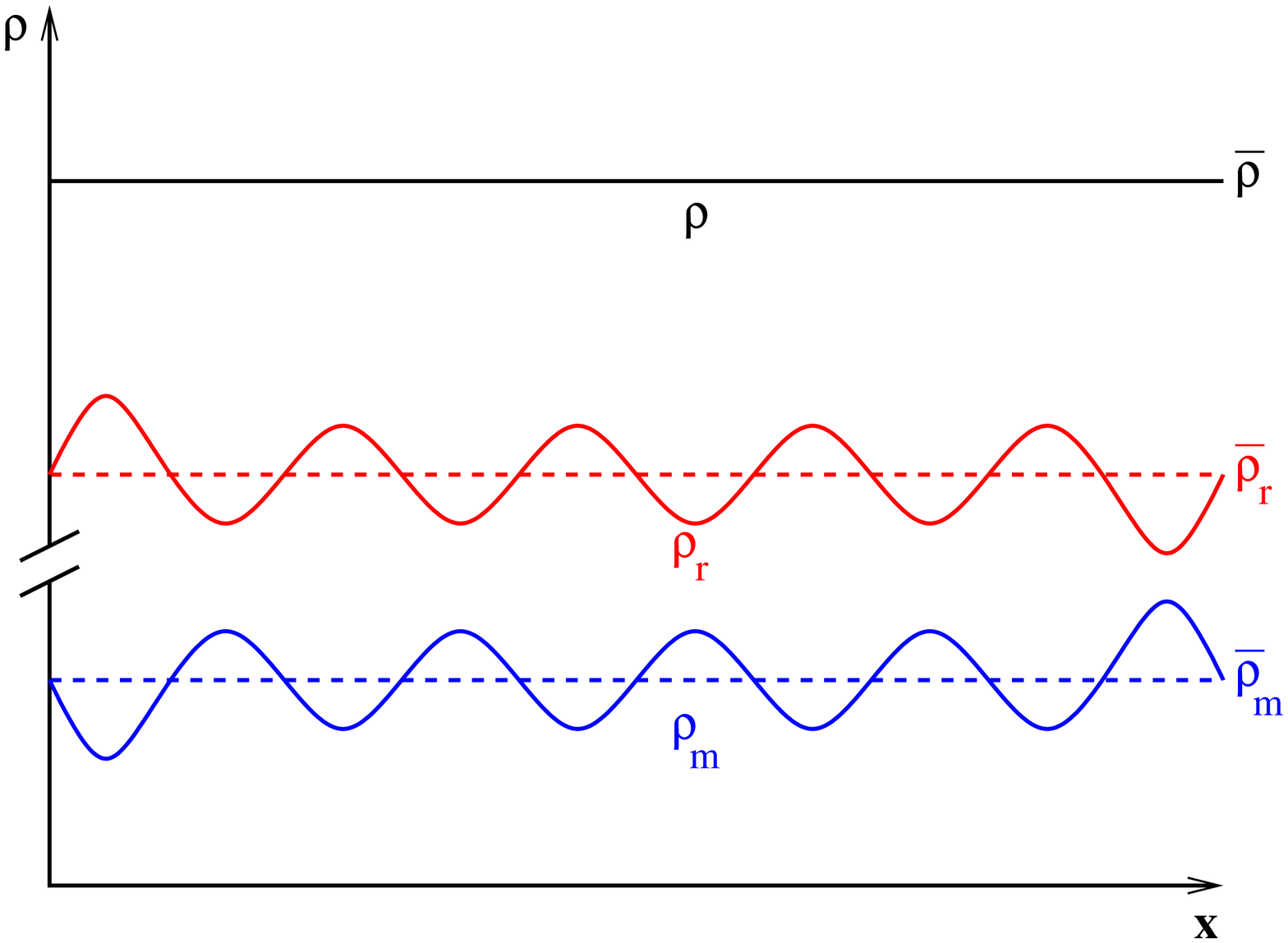}
}
\end{center}
\caption{(a) History of fluctuations. 
The beginning of radiation dominated era is $t_{\rm rad}$.
(b) An example of adiabatic initial fluctuations. (c) An example
of isocurvature initial fluctuations at time $t_{\rm rad}$.}
\end{figure}


The most studied possibility is pure adiabatic initial mode. Then there is
no entropy perturbation at $t_{\rm rad}$,
$ S_{\rm rad} \equiv  S_{c\gamma} = \frac{\delta\rho_{c}}{\bar{\rho}_{c}} -
\frac{3}{4}\frac{\delta\rho_{\gamma}}{\bar{\rho}_{\gamma}} = 0$,
but the total energy density fluctuates or more precisely there is
a spatial perturbation in the comoving curvature
$\langle |\mathcal R|^2_{\rm rad} \rangle \neq 0$. In Fig.~1(b)
I have an example, where the spatial fluctuations
in matter and radiation energy density are in the same phase 
yielding to an initial fluctuation in the total energy density.

In the isocurvature case the specific entropy fluctuates spatially
$ S_{\rm rad} \neq 0$,
but there is no initial fluctuations in the total energy density or more
precisely in the comoving curvature, $\langle |{\cal R}|^2 \rangle = 0$.
For example, the fluctuation in matter and radiation could cancel each
other giving spatially constant total energy density as in Fig~1(c).


The evolution of perturbations is described by second order
differential equations, adiabatic and isocurvature initial conditions
being two independent modes. Hence the most general initial condition
is a mixture of adiabatic and isocurvature fluctuations. The evolution
equations tell how adiabatic and isocurvature initial fluctuations are
converted into the temperature fluctuation present at the last scattering
surface and finally into the presently observable temperature
(or polarization) anisotropy described by the angular power spectrum
that contains a series of peaks and valleys. If the other cosmological
parameters are kept fixed, but one changes form adiabatic initial conditions
to the isocurvature ones, then the resulting angular power spectra are roughly
in the opposite phases as can be seen in Fig.~2(a).

Already the first data sets by Boomerang and Maxima could be used to constrain
uncorrelated mixture of adiabatic and isocurvature fluctuations in flat
universe models. We found the maximum $2\sigma$ allowed isocurvature
contribution to the quadrupole ($l=2$) temperature anisotropy to be 56\% 
and to the first acoustic peak ($l \sim 200$) about 13\% \cite{Enqvist:2000hp},
see Fig.~2(b). Since in open universe all the features of the angular
power spectrum are shifted towards right (larger $l$, smaller scales)
and in closed universe towards left, it still seemed to be possible
to fit the first acoustic peak by open or closed pure isocurvature models.
However, the second data releases by Boomerang and Maxima  \cite{Netterfield:2001yq}
identified also the second acoustic peak so well that the
``adiabatic peak structure'' became evident and we could finally rule
out all pure CDM isocurvature models \cite{Enqvist:2001fu}. From Fig.~2(c)
one can see even by an eye that the pure isocurvature does very badly with
the data. Nevertheless, {\em mixed} correlated or uncorrelated models remain as
an interesting possibility.

\section{Correlation and the WMAP}

The initial fluctuations for the CMB physicist
are the comoving curvature perturbation $\hat{\cal R}_{{\rm rad}}(k)$
and the entropy perturbation $\hat{\cal S}_{{\rm rad}}(k)$ at the
beginning of the radiation dominated era. By hats I remind you that
these are Gaussian random variables. They are related to the
fluctuations at the horizon exit ($t_*$) during inflation by \cite{Gordon:2000hv}
\begin{equation}
  \begin{pmatrix}
    \hat{\cal R}_{{\rm rad}}(k) \\ 
    \hat{\cal S}_{{\rm rad}}(k)
  \end{pmatrix}
  =
  \begin{pmatrix}
    1 & T_{{\cal R}{\cal S}}(k) \\ 
    0 & T_{{\cal S}{\cal S}}(k)
  \end{pmatrix}
  \begin{pmatrix}
    \hat{\cal R}_*(k) \\ 
    \hat{\cal S}_*(k)
  \end{pmatrix}\,,
\end{equation}
where $T_{{\cal R}{\cal S}}(k)$ and
$T_{{\cal S}{\cal S}}(k)$ are the transfer functions that describe
the evolution of perturbations form $t_*$ to $t_{\rm rad}$.
In most cases, they
are found only numerically by solving the evolution equations and
modeling the reheating process.

The initial perturbations  $\hat{\cal R}_{{\rm rad}}(k)$
and $\hat{\cal S}_{{\rm rad}}(k)$ are usually approximated 
by power laws \cite{Gordon:2001ph,Peiris:2003ff},
\begin{equation}
  \hat{\mathcal{R}}_{\mathrm{rad}} =
  A_{r}\Bigr(\textstyle\frac{k}{k_{0}}\Bigl)^{n_{1}}\hat{a}_{r}(\mathbf{k}) +
  A_{s}\Bigr(\textstyle\frac{k}{k_{0}}\Bigl)^{n_{3}}\hat{a}_{s}(\mathbf{k}),
\ \ \mbox{ and } \ \ 
  \hat{\mathcal{S}}_{\mathrm{rad}} =
  B\Bigr(\textstyle\frac{k}{k_{0}}\Bigl)^{n_{2}}\hat{a}_{s}(\mathbf{k}),
\label{jvaliviita:initialfluct}
\end{equation}
which is a good approximation assuming that
``everything'' changes slowly during inflation. Actually,
e.g., in some cases of double inflation this is not true \cite{Tsujikawa:2002qx}.
In an ordinary pure adiabatic case one would have
only the first term
$A_{r}(k/k_{0})^{n_{1}}\hat{a}_{r}(\mathbf{k})$.
Now I have additional terms coming from the entropy perturbation
during inflation. In my terminology
$A_{r}(k/k_{0})^{n_{1}}\hat{a}_{r}(\mathbf{k})$
is called the first adiabatic component, 
$A_{s}(k/k_{0})^{n_{3}}\hat{a}_{s}(\mathbf{k})$ the
second adiabatic component (generated by the entropy
perturbation during inflation if the trajectory in the
multi field space is curved \cite{Gordon:2000hv,Gordon:2001ph}), and
$ B(k/k_{0})^{n_{2}}\hat{a}_{s}(\mathbf{k})$
the isocurvature component.
$A_{r}$, $A_{s}$, and $B$ are amplitudes and $n_i$ spectral indices.
Moreover, $\hat a_r$ and $\hat a_s$ are Gaussian random variables obeying
$\langle \hat a_r \rangle = 0$, $\langle \hat a_s \rangle = 0$, and
$\langle \hat a_r(\mathbf{k}) \hat a_s^*(\mathbf{k}')\rangle  
= \delta_{rs}\delta^{(3)}(\mathbf{k} - \mathbf{k}')\,$.
\begin{figure}[t]
\begin{center}
{\bf (a)}
\includegraphics[width=0.18\textwidth]{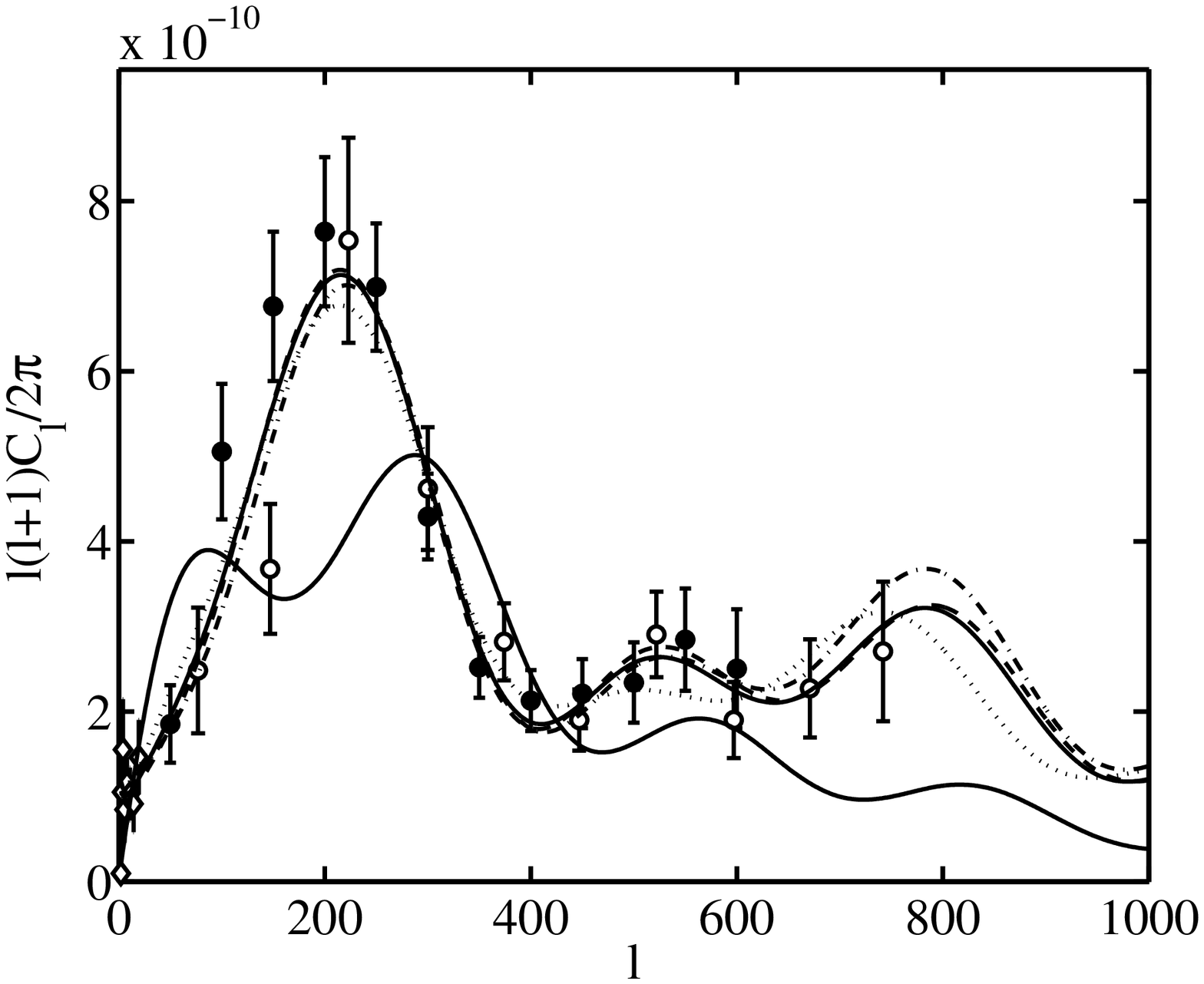}
\hspace{0.5cm}
{\bf (b)}
\includegraphics[width=0.18\textwidth]{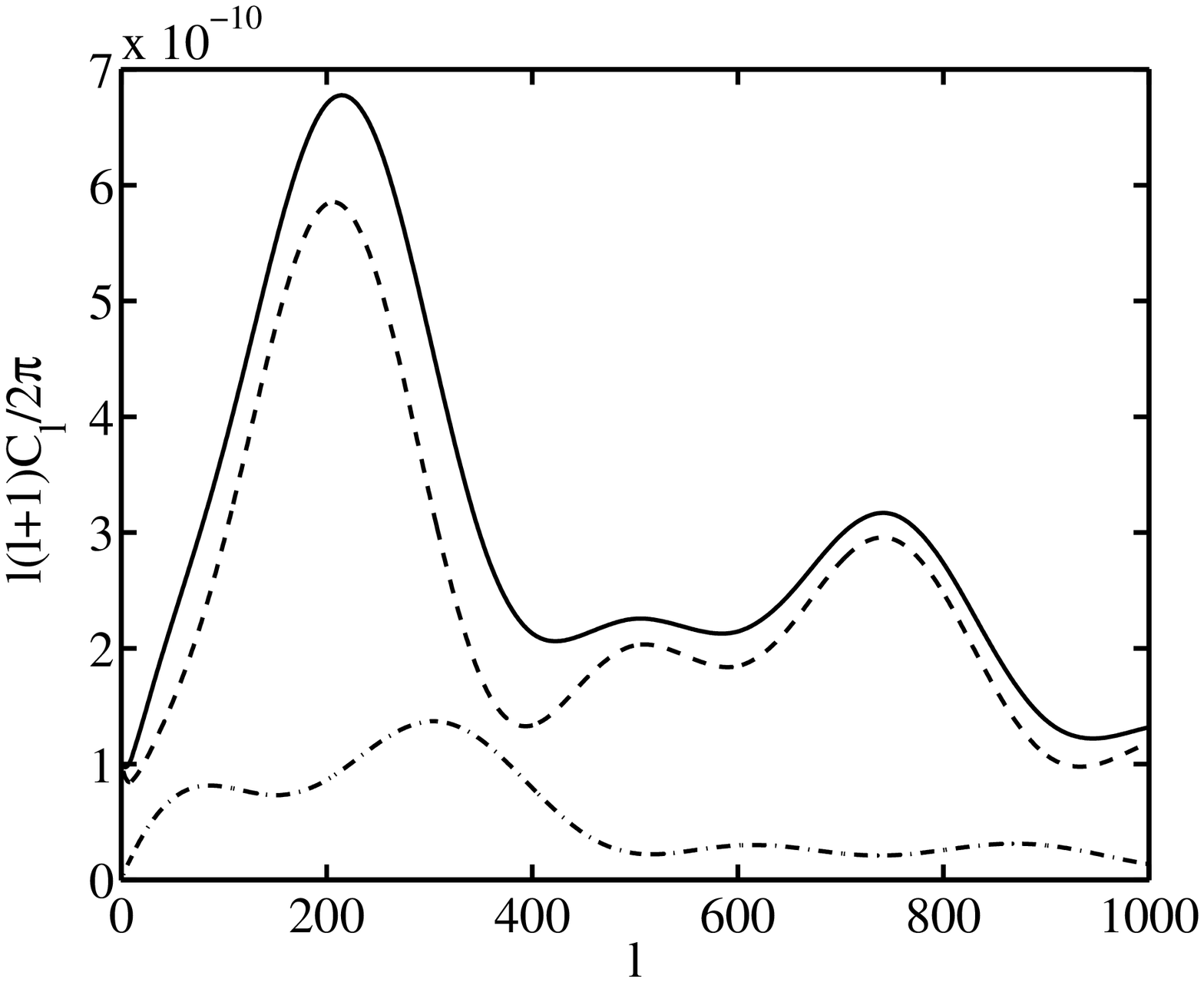}
\hspace{0.5cm}
{\bf (c)}
\includegraphics[width=0.18\textwidth]{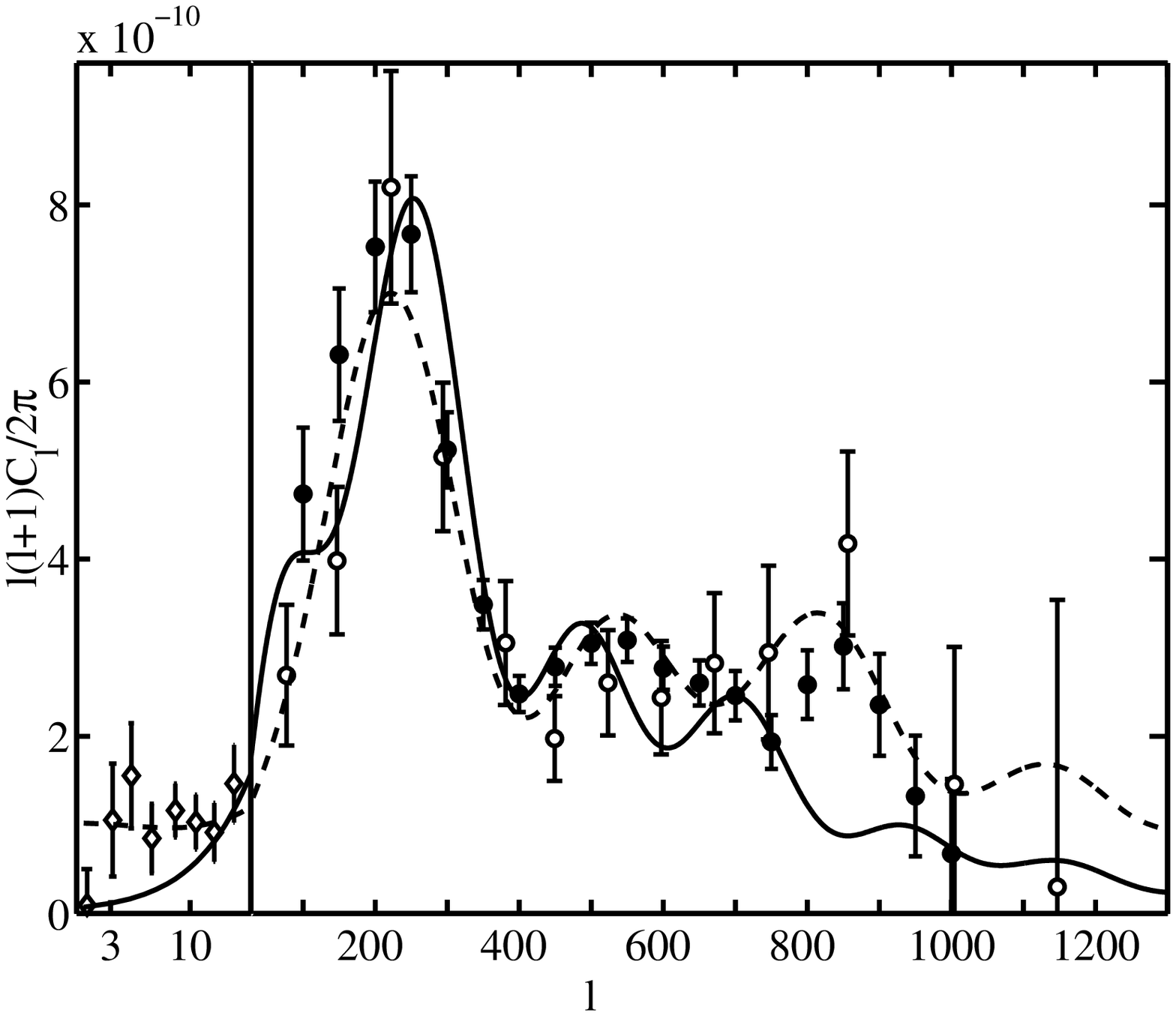}
\end{center}
\caption{(a) The model with peak at $l \sim 300$ is an isocurvature model while
the other models are dominated by adiabatic fluctuations.
(b) The maximum $2\sigma$ allowed isocurvature contribution (in flat $\Omega = 1$
universe) to
$l = 200$. Dot dashed line is the isocurvature contribution, dashed line
the adiabatic contribution and solid line the total temperature angular power.
(c) The best-fit closed universe 
pure isocurvature model (solid line) compared with a typical
adiabatic model (dashed line).
Figure taken from\protect\cite{Enqvist:2001fu}.%
}
\end{figure}

Now there is a 100 \% correlation between the second abiabatic and
the isocurvature component,
$\langle \hat{\mathcal R}(\mathbf{k}) \hat{\mathcal S}^{\ast}(\mathbf{k}')\rangle
\vert_{\mathrm{rad}} = A_s B (\textstyle k/k_{0})^{n_3 + n_2}
\delta^{(3)}(\mathbf{k} - \mathbf{k}')$. This can be parametrized by
some overall amplitude (at the reference scale $k_0$) $A^2 = A_r^2 + A_s^2$,
isocurvature fraction $f_{\rm iso} = |B/A|$, and 
correlation amplitude $\cos\Delta = \mbox{sign}(B) A_s/A$
as follows: 
$\langle \hat{\mathcal R}(\mathbf{k}) \hat{\mathcal S}^{\ast}(\mathbf{k}')\rangle
=
 A^2 f_{\rm iso} \cos\Delta  (\textstyle k/k_{0})^{n_3 + n_2}$
$\delta^{(3)}(\mathbf{k} - \mathbf{k}')$.
The final angular power spectrum will be a combination of four
different components: the first adiabatic, the second adiabatic,
the isocurvature and the correlation between the second adiabatic
and the isocurvature component
\begin{equation}
  C_{l} = A^{2}\bigr[ \sin^{2}(\Delta) C_{l}^{\mathrm{ad1}}
 +  \cos^{2}(\Delta) C_{l}^{\mathrm{ad2}}
  + f_{\mathrm{iso}}^{2} C_{l}^{\mathrm{iso}} + 
 f_{\mathrm{iso}} \cos(\Delta)
  C_{l}^{\mathrm{cor}} \bigl].
\label{jvaliviita:finalcl}
\end{equation}
W{\textsc map} team already analyzed this kind of models \cite{Peiris:2003ff},
but for simplicity, they assumed the two adiabatic spectral indices to be equal,
$n_1 = n_3$. This is not theoretically well motivated, since $n_1$ comes
from the curvature perturbation and $n_3$ from the entropy perturbation.
If one really needs to simplify the analysis, one should consider
to put $n_3$ and $n_2$ equal.

In our Letter \cite{Valiviita:2003ty}
we allow for $n_1$, $n_2$, and $n_3$ to vary independently. To match
the historical notation we redefine the spectral indices so that
the value 1 corresponds to the scale free case: 
$n_{\rm ad1} - 1  =  2n_1$,
$n_{\rm ad2} - 1  =  2n_3$, and
$n_{\rm iso} - 1  =  2n_2$. We use a grid method
to scan the parameter space of 9 parameters
($\sim10^{10}$ combinations):
$\tau$, $\Omega_{\Lambda}$, $\omega_{b}$, $\omega_{c}$,
$n_{\mathrm{ad1}}$, $n_{\mathrm{ad2}}$, $n_{\mathrm{iso}}$,
$f_{\mathrm{iso}}$, and $\cos(\Delta)$.
Marginalization by integration is adopted,
since some likelihoods are non-Gaussian.


Fig.~3(a) shows our main result that the data do allow unequal adiabatic
spectral indices, and actually, favour models where the two
adiabatic components have opposite spectral tilts  ($n_{\mathrm{ad1}} > 1$
and $n_{\mathrm{ad2}} < 1$ or vice versa). This leads
to a running effective adiabatic spectral index. The adiabatic
initial power is 
$ {\cal P_{R}}(k) = A^2\left[ \sin^2\Delta
(\textstyle{k/k_0})^{n_{\rm ad1}-1}
+ \cos^2\Delta (\textstyle{k/k_0})^{n_{\rm ad2} - 1} \right]$,
see equation (\ref{jvaliviita:initialfluct}) and the definitions above equation
(\ref{jvaliviita:finalcl}). Writing this as a single power law
$
  {\cal P_{R}}(k) = D (\textstyle{k/k_0})^{n_{\rm ad} - 1}\,,
$
where $D$ is an amplitude and
${n_{\rm ad} - 1} =
d\ln{\cal  P_{R}}(k)/d\ln k
$, one always gets non-negative first derivative for the common
adiabatic spectral index $\nadi$;
\begin{equation}
  \frac{dn_{\rm ad}(k)}{d\ln k}
  = \frac{\sin^2\Delta\cos^2\Delta
(n_{\rm ad1} - n_{\rm ad2})^2 k^{n_{\rm ad1}+n_{\rm ad2}}}{[\sin^2\Delta
    k^{n_{\rm ad1}} + \cos^2\Delta k^{n_{\rm ad2}}]^2}.
\end{equation}

Using the \wmap data only we get a $2\sigma$ upper bound for the
isocurvature fraction, $f_{\rm iso} < 0.84$. Since the angular power
alone cannot give ``any'' upper bound for the isocurvature spectral
index, we have to put some prior for it; $n_{\rm iso} < 1.84$.
We assume from the \wmap team analysis 
that the large scale structure data (e.g., 2dF) would
give about $n_{\rm iso} < 1.80$ motivating our prior. In any case,
we can compare our result to the result which we would get with
the \wmap team restriction $n_{\mathrm{ad2}} = n_{\mathrm{ad1}}$.
This simplification leads to unrealistically strict
upper bound, $f_{\rm iso} < 0.74$. 

In Figs.~3(b)\&(c) I have an example of a $2\sigma$ allowed
correlated model.
In the case of negative correlation, as here, the isocurvature
contribution to the lowest multipoles $l$ in TT spectrum 
can be quite large due to cancellation by the correlation part.
In TE spectrum the cancellation is not as exact so that there
one gets enough power at the lowest multipoles. Actually, in 
this example, the isocurvature component dominates TE spectrum
at the quadrupole $l=2$. That is why even quite a small optical
depth due to reionization ($\tau$) is enough to give a reasonable
total power at the quadrupole. Thus in correlated models the measured
high quadrupole in TE spectrum can be achieved without having
as large $\tau$ as reported by the \wmap team. So one should
keep in mind a possible degeneration between the optical depth and
correlation  contribution. On the other hand, a positive correlation
could nicely lead to even more power in TE at low $l$, but unfortunately
it tends to give a high quadrupole also in TT, which is not favoured by
the data. An interesting possibility to explain the low quadrupole
in TT would be to have a very strong negative correlation which 
``eats'' some power from TT quadrupole. Then in TE this could be
compensated by a large $\tau$. Suitably strong negative
correlation contribution to the total TT power requires that the
second adiabatic and isocurvature components dominate over the first
adiabatic component.

\begin{figure}
\begin{center}
{\bf(a)}
   \includegraphics[width=0.205\textwidth]{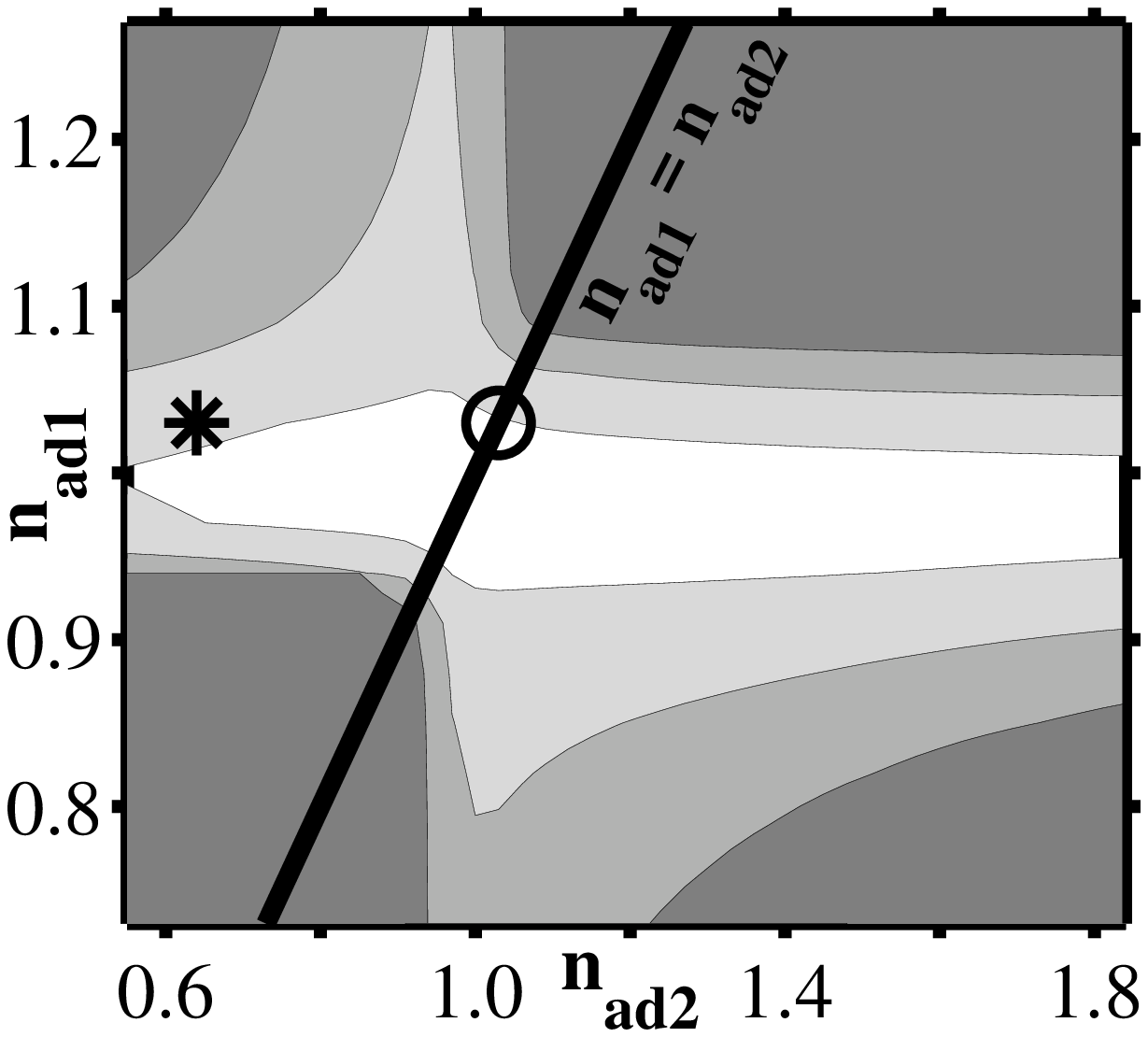}
\hspace{0.5cm}
{\bf (b)}
   \includegraphics[width=0.225\textwidth]{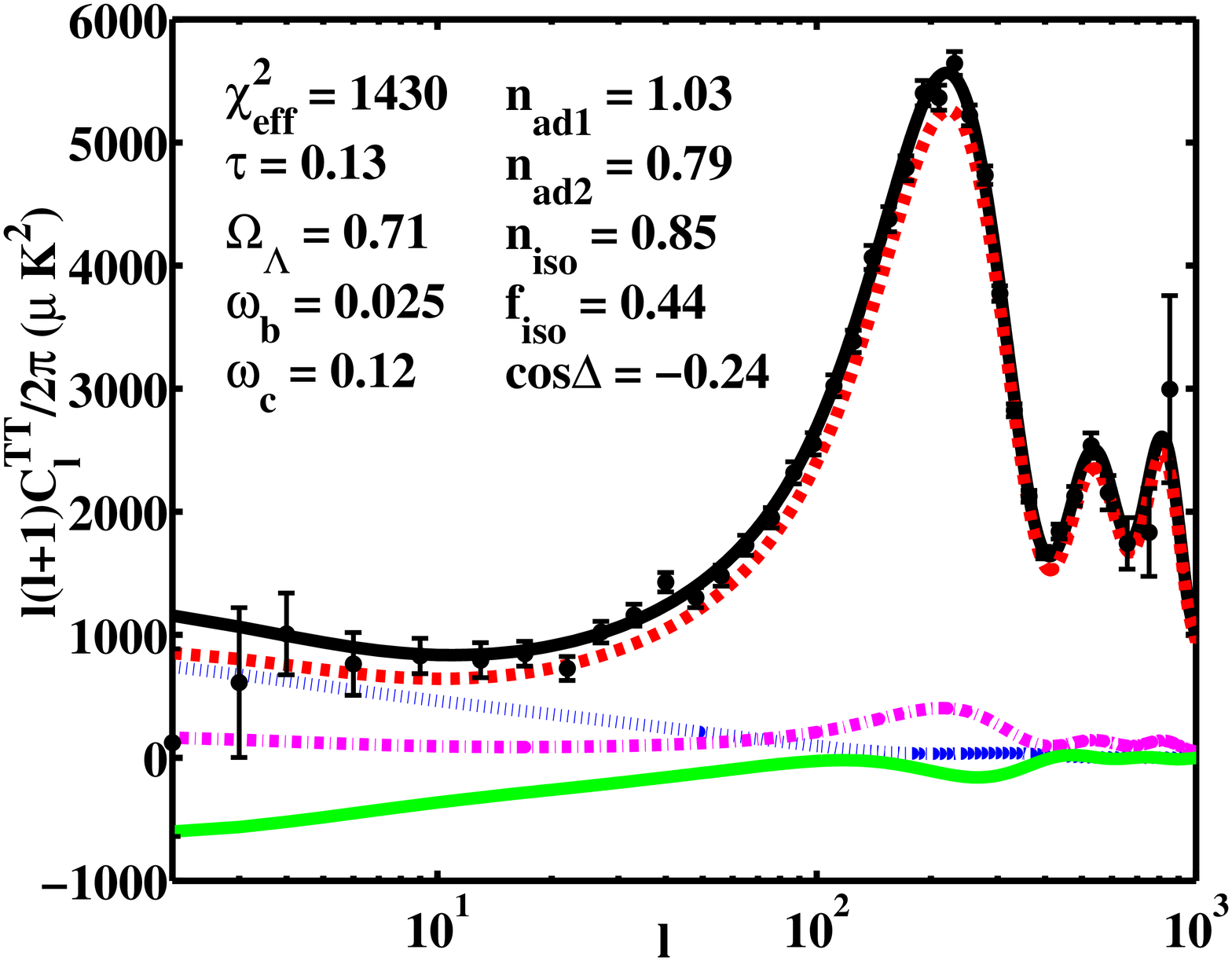}
\hspace{0.5cm}
{\bf (c)}
   \includegraphics[width=0.225\textwidth]{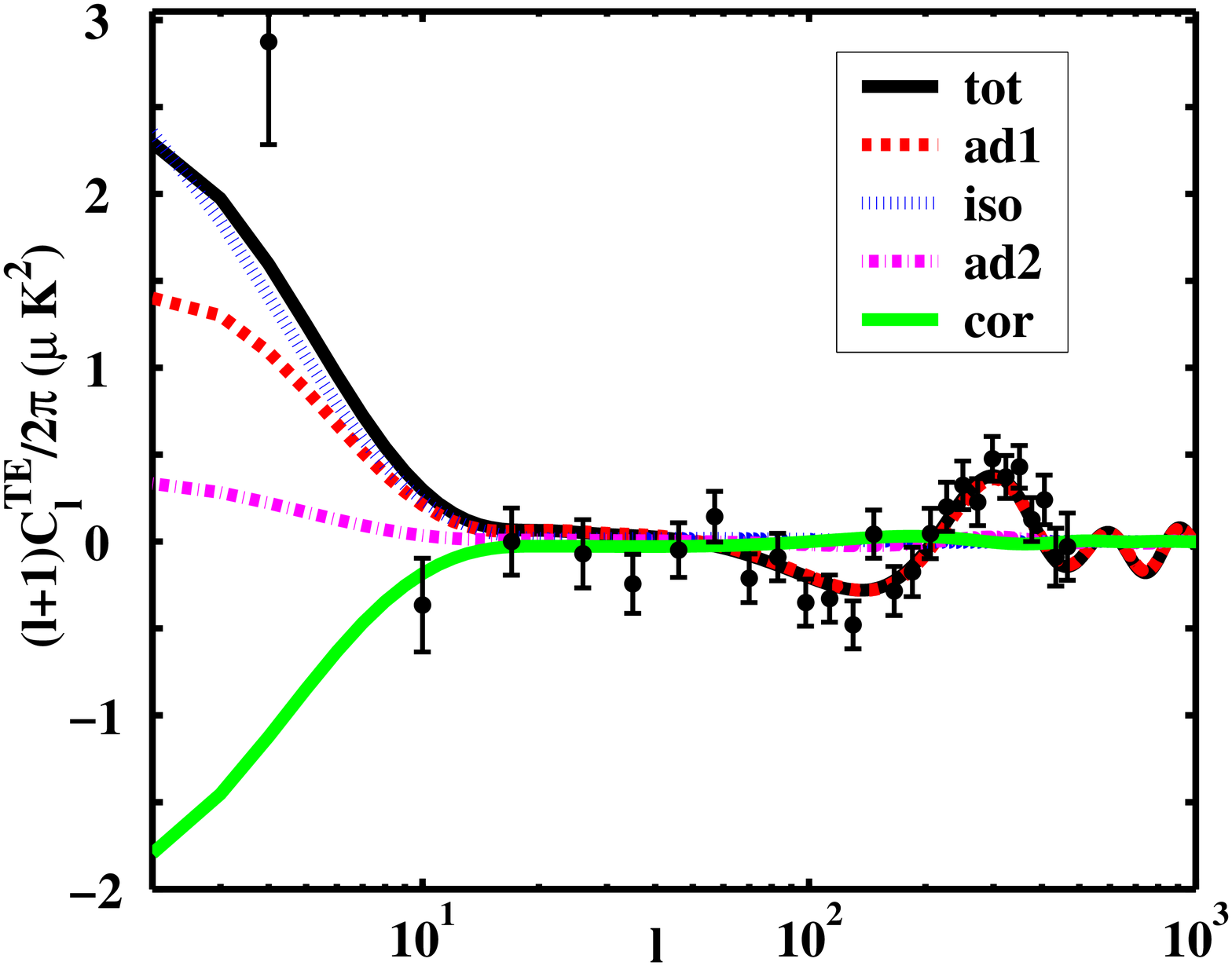}
\end{center}
\caption{(a) The 68.3\%/$1\sigma$ (white),
    95.4\%/$2\sigma$ (light gray),
    99.7\%/$3\sigma$ (medium gray),
    and more than $3\sigma$ (dark gray)
    confidence levels for our general models.
    (b)\&(c) Angular spectra
    of a 2$\sigma$ allowed example model.}
\end{figure}



\end{document}